# Effect of synthesis temperature on superconducting properties of *n*-SiC added bulk $MgB_2$ superconductor

Arpita Vajpayee[1,*], R. Jha[1], A. K. Srivastava[1], H. Kishan[1], M. Tropeano[2], C. Ferdeghini[2] and V.P.S. Awana[1]

[1]*Quantum Phenomenon and Applications (QPA) Division, National Physical Laboratory (CSIR), New Delhi-12, India*

[2]*CNR-Spin, corso Perrone 24, 16152 Genova, Italy*

We study the effect of synthesis temperature on the phase formation in *nano*(*n*)-SiC added bulk $MgB_2$ superconductor. In particular we study: lattice parameters, amount of carbon (C) substitution, microstructure, critical temperature ($T_c$), irreversibility field ($H_{irr}$), critical current density ($J_c$), upper critical field ($H_{c2}$) and flux pinning. Samples of $MgB_2+(n\text{-SiC})_x$ with x=0.0, 0.05 & 0.10 were prepared at four different synthesis temperatures i.e. 850, 800, 750, and 700$^o$C with the same heating rate as 10$^o$C/min. We found 750$^o$C as the optimal synthesis temperature for *n*-SiC doping in bulk $MgB_2$ in order to get the best superconducting performance in terms of $J_c$, $H_{c2}$ and $H_{irr}$. Carbon (C) substitution enhances the $H_{c2}$ while the low temperature synthesis is responsible for the improvement in $J_c$ due to the smaller grain size, defects and *nano*-inclusion induced by C incorporation into $MgB_2$ matrix, which is corroborated by elaborative *HRTEM* (high-resolution transmission electron microscopy) results. We optimized the the $T_c(R=0)$ of above 15K for the studied *n*-SiC doped and 750 $^0$C synthesized $MgB_2$ under 140 KOe field, which is one of the highest values yet obtained for variously processed and *nano*-particle added $MgB_2$ in literature to our knowledge.

[*]Corresponding Author
e-mail-vajpaya@mail.nplindia.ernet.in
Fax No. 0091-11-45609310: Phone No. 0091-11-45609210

Introduction

The discovery of the superconductivity in MgB$_2$ with $T_c$ = 39 K [1] stimulated much attention from fundamental to applied research. Due to its relatively high critical temperature, low cost and transparency of the grain boundaries to the current flow [2], MgB$_2$ superconductor is one of the best candidates for engineering applications in the temperature range 20-30 K and can be competitive with traditional Nb based superconductors and with the so called second generation HTS wires. However, its critical current density ($J_c$) is suppressed under magnetic fields due to low upper critical fields ($H_{c2}$) and poor flux pinning in the clean material. Therefore, improvement of $J_c$ in magnetic field is needed for applications in extensive application fields. There have been several demonstrations of enhanced vortex pinning and critical current density, $J_c$, in the magnesium diboride by partial substitution of boron for carbon, trough doping with SiC, B$_4$C, carbon nano tube (CNT) and nano-C doping [3-11]. Substitution of C in the B site also significantly increases the upper critical field, $H_{c2}$ [12-14]. However, the resulting pinning effects by doping were not quantitatively consistent among many research groups [15]. Since superconducting properties of MgB$_2$ are very sensitive to phase purity, size of starting boron powder and synthesis conditions so the flux pinning properties of the samples varied effectively in each paper [16].

Some of us previously found that a significant flux pinning enhancement in MgB$_2$ could be easily achieved using *nano* SiC as an additive [3]. The Si and C released from the decomposition of *nano* SiC at the time of formation of MgB$_2$ formed Mg$_2$Si and substituted at B sites respectively. The C substitution for B resulted in a large number of intra-granular dislocations and dispersed nano-size impurities, which are both responsible for the significant enhancement in flux pinning. However, other micro-structural factors such as grain size or connectivity of MgB$_2$ grains, which determine the flux pinning properties of undoped MgB$_2$, have not been well clarified as functions of synthesis temperature. Based on these backgrounds, here, we study the effect of synthesis temperature on phase formation, critical temperature ($T_c$), upper critical field ($H_{c2}$), critical density ($J_c$), irreversibility field ($H_{irr}$) and microstructures of doped MgB$_2$.

Experimental details

Our polycrystalline MgB$_2$+$n$-SiC$_x$ (x = 0, 5-wt% & 10-wt%) samples were synthesized by solid-state reaction route. The Mg powder used is from *Reidel-de-Haen*, amorphous B powder is from *Fluka* (of assay 95-97%) and *n*-SiC powder is from *Aldrich* (<100 nm). For synthesis the samples, the stoichiometric amounts of ingredients were ground thoroughly, palletized using hydraulic press and put in a tubular furnace for 2.5 hours in argon atmosphere at four different temperatures: 700°C, 750°C, 800°C & 850°C. Finally the furnace cooled in the same atmosphere of argon to room temperature.

Constituent phases of the samples were analyzed by the *Rigaku miniflex II* powder x-ray diffractometer using *CuK$_\alpha$* radiation. The grain morphology and microstructures were examined by high-resolution transmission electron microscopy (*HRTEM*) (model *Tecnai G2 F30 STWIN* field emission gun supported, operated at the electron accelerating voltage of 300 kV). The magneto-resistivity was measured with *H* applied perpendicular to current direction, using four-probe technique. The magnetization measurements were carried out on *Quantum Design* Physical Property Measurement System (*PPMS*) system. Critical current density ($J_c$) was determined from magnetization loop using the extended Bean model.

Results and Discussion

**(A) Phase formations and crystallographic interpretations by powder XRD:**

In order to clarify the influence of synthesis temperature ($T_s$) on the phase formation of MgB$_2$, X-ray diffraction (XRD) pattern were recorded at room temperature. Figure 1(a) shows the XRD patterns for all the *n*-SiC doped samples along with pristine MgB$_2$ synthesized at 700°C, 750°C, 800°C & 850°C. In case of pure MgB$_2$ all characteristic peaks are obtained and their respective indexing is shown in the figure itself. The structure of MgB$_2$ belongs to space group *P6/mmm*. Powder XRD analysis revealed that nearly single-phase MgB$_2$ bulks containing only a minor fraction of MgO (marked by 'o' in figure) were obtained for all the samples heated at above 700°C, while the samples heated at 700°C showed strong diffraction peaks due to unreacted magnesium (marked by '$' in the Figure 1(a)) in the sample. The volume fraction of

unreacted Mg is nearly 25% in pure $MgB_2$ sample synthesized at $700^oC$. MgO fraction is almost the same in all the samples irrespective of SiC content and synthesis temperature.

Samples doped with SiC showed a considerable amount of $Mg_2Si$ (marked with * in the figures) and minor quantities of unreacted SiC (+). In particular for both $MgB_2$+n-$SiC_x$, the {hkl} planes {111}, {200}, {220}, {311} and {400} of the $Mg_2Si$ are noticed clearly in Figure1. At x = 5-wt% doping level the samples consist of a major phase of $MgB_2$ with minority phase of $Mg_2Si$; increasing the doping level of *n*-SiC at x = 10-wt%, the amount of this non-superconducting phase increases. The formation of $Mg_2Si$ in SiC-doped samples indicates the dissociation of the SiC and the reaction of Si with Mg (SiC starts to react with Mg at a temperature as low as $600^oC$). In fact, this decomposition of SiC and the formation of the $Mg_2Si$ phase are reported in almost all the SiC doping studies [3,4,6,17]. No other impurity phases like $MgB_4$, $MgB_7$, $Mg_2C_3$ and $MgB_2C_2$ are detected in any one of the samples.

As far as the majority $MgB_2$ phase is concerned, the peak {100} situated between $2\theta = 33^o$ and $2\theta = 34^o$ shifts towards the higher angle with increasing doping content of SiC, indicating the contraction in a-axis in crystal lattice. The lattice parameters, *a* and *c*, of the hexagonal $AlB_2$ type structure of $MgB_2$ are calculated using these peak shifts, and their variation is reported in Table 1. The change in *c* with increasing x is relatively small as compared to *a* parameter. The *a*-axis parameter shows a larger drop with x for all the temperatures. The *a* decrease indicated the effective partial substitution of B by C [3,4,5-13,17-18]. The substituent C atoms are readily available from the SiC. The actual C substitution level for our $MgB_2$ + $(SiC)_x$ samples can be estimated indirectly from the change in lattice parameters using formula x = 7.5 $\Delta(c/a)$ where $\Delta(c/a)$ is the change in *c/a* compared to pure sample [18,19]. Figure 1(b) depicts (from the bottom to the top) the variation of *a*-axis lattice parameter, *c/a* value, actual C substitution level (x) in $Mg(B_{1-x}C_x)_2$ and *FWHM* of (110) peak with synthesis temperature. Figure 1(b) shows an increase in c/a value at a particular synthesis temperature as we add the *n*-SiC in $MgB_2$; which clearly indicates the presence of the lattice strain in doped samples. The lattice strain can be attributed to the C substitution in the structure and unreacted C inside the grains [20]. It is observed that *c/a* and x values are change moderately for 750, 800 and $850^oC$ samples in comparison to $700^oC$ samples. Except $700^oC$, the low temperature synthesis causes larger values of *FWHM*, suggesting smaller grain size and imperfect crystallinity. Figure 1(b) revealed that change in *a*-axis parameter, *c/a* values, *FWHM* values and actual C substitution level for 700°C samples are lower than those of other samples, which are heated at relatively higher temperature; hence it seems that 700°C synthesis temperature is comparatively lower temperature for partial substitution of C at B site in $MgB_2$ matrix.

**(B) DC and AC susceptibility measurements:**

Figure 2(a) depicts the DC susceptibility versus temperature $\chi(T)$ plots in FC and ZFC case for an applied field of 1 mT for the series of samples prepared at 750 °C. It is evident from the figure that pure $MgB_2$ undergoes a sharp superconducting transition with an onset at 38.6 K within 1 K temperature interval. All the samples exhibit one-step transition from normal state to superconducting state, however the transition width increases a bit by increasing the amount of dopant.

The temperature dependence of the real ($X'$) and the imaginary ($X''$) parts of the AC susceptibility is illustrated in Figure 2(b) for all the three samples realized at 750 °C. For these AC susceptibility measurements, an AC magnetic field of 1 mT rms value with frequency $f = 33$ Hz has been used in the absence of a DC field. Below the critical temperature, a sharp decrease in the real part of the AC susceptibility occurs, which reflects the diamagnetic shielding. In addition, below $T_c$ a peak appears in $X''$, reflecting losses related to the flux penetrating inside the grains. No evidence of a two peaks behavior (granularity) was detected. As the SiC content increases, the onset of diamagnetic shielding in $X'$ and the peak position in $X''$ shift towards lower temperatures.

Figure 3 shows the variation of transition temperature ($T_c$) with synthesis temperature ($T_s$) for the undoped and doped samples. $T_c$ is deduced from the onset temperature of the diamagnetic transition ($T_c^{dia}$) in the DC susceptibility measurements. For the undoped sample $T_c$ shows an increase with increasing preparation temperature, which can be attributed to the improvement of crystallinity due to higher temperatures. The $T_c$ for the doped samples depends also on the C substitution level:the competitive behavior of these two factors produced the curves of Figure 3.

## (C) Irreversibility field ($H_{irr}$) and critical current density ($J_c$):

The magnetic hysteresis loops for the undoped and doped samples prepared at 750°C are shown in Figure 4 at $T = 5, 10, 20$ K and under up to 14 Tesla applied field. This figure clearly demonstrates that at $T = 5$ K the closing of $M(H)$ loop for pure sample is at ~9 Tesla, that become ~12 Tesla for x=5-wt% *n*-SiC doped sample and $M(H)$ loop is still open at 14 Tesla for x=10. This indicates that there is quite significant improvement in irreversibility field values by addition of *n*-SiC in parent compound at the reaction temperature of 750°C. The irreversibility fields ($H_{irr}$) are derived from the fields at which the magnetic hysteresis loop gets nearly closed; with the criterion of giving the $J_c = 100$ A/cm$^2$. To know the effect of synthesis temperature along with doping level of *n*-SiC on $H_{irr}$ values a plot is drawn in $H_{irr}$ versus $T_s$ and it is shown in Figure 5(a) for all the three samples [(i) undoped, (ii) 5-wt% & (iii) 10-wt% *n*-SiC doped] at 5 K, 10 K & 20 K. Careful look of Figure 5(a) implies that doping with *n*-SiC has significantly improved the irreversibility field ($H_{irr}$) for all the synthesis temperatures. Among all the synthesis temperatures the 750°C gives the best value of $H_{irr}$, which can be attributed to grain boundary pinning due to smaller grain size at low synthesis temperature. Therefore, we plotted the variation of $\mu_0 H_{irr}$ with *n*-SiC concentration for 750°C samples separately in Figure 5(b). The 10-wt% n-SiC doped samples the highest values of $H_{irr}$ at $T_s$=750°C. The values of $\mu_0 H_{irr}$ for the 10-wt% *n*-SiC added sample reached 14, 11.8 & 6.5 Tesla, compared to 8.9, 7.9, 5.0 Tesla for the pure one at 5, 10 & 20 K respectively. The spectacular enhancement in $\mu_0 H_{irr}$ values, being seen in Figures 5(b) is definitely due to improvement in flux pinning in MgB$_2$ by the *n*-SiC doping and low temperature synthesis.

The magnetic $J_c$ for all the samples was calculated using Bean's model from the $M(H)$ loop at 5, 10 & 20 K. Figure 6(a) shows the magnetic $J_c$ vs $\mu_0 H$ for all the samples for all the four reaction temperatures at 20 & 5 K. We found from the XRD and phase analysis that the 700°C samples are not very much phase pure and also there is no appreciable values of actual C substitution level in the doped samples at this reaction temperature. Accordingly the first figure of Figure 6(a) also represents a very less improvement in 5-wt% doped samples in comparison to pure MgB$_2$ sample. For the other temperature reaction, among all the doped samples 10-wt% *n*-SiC doped sample gives the best $J_c(H)$ performance. The samples reacted at 800 and 850 °C show the almost similar behavior of $J_c(H)$ performance. In these two series of samples, for undoped $MgB_2$, $J_c$ drops rapidly in the presence of magnetic field and for both is almost negligible above 4 Tesla and around 7 Tesla at 20 K and 5 K respectively; for the 10-wt% n-SiC doped samples the $J_c$ becomes negligible at around 5 and 11 Tesla respectively. Among all the four *series* the one at 750 °C demonstrates the superior $J_c(H)$ performance. The $J_c$ is 16 times higher than pure one in case of 5-wt% n-SiC doped sample and 30 times higher for the 10-wt% n-SiC doped sample at 5 K in 8.5 Tesla field. The $J_c$ values are $1.1 \times 10^2$ A/cm$^2$ and $1.2 \times 10^3$ A/cm$^2$ at 5 K in the high field of 12 T for 5-wt% and 10-wt% n-SiC doped samples respectively.

Figure 6(b) shows the dependence of magnetic $J_c$ on synthesis temperature at 5 K and 6 Tesla for the undoped one and at 5 K and 9 Tesla for the doped samples. Both pure and SiC doped have the best $J_c$ for lower synthesis temperature. The $J_c$ values decrease gradually with increasing synthesis temperature for the SiC doped and undoped samples. These $J_c$ values are quite competitive with reported literature [5, 21-22] and little bit higher than in comparison to Ref. [16, 23-24].

The observation of dependence of $J_c$ on SiC concentration and on synthesis temperature strongly suggests that the following two parameters affect the $J_c$: i) enhancement of grain boundary pinning by growing small crystals at lower synthesis temperature and ii) introduction of defects/dislocations by C substitution. We observed that lower synthesis temperature i.e. 750°C is best for the better $J_c(H)$ performance for undoped $MgB_2$ as well as for the n-SiC doped $MgB_2$. As opposed to n-C doping, n-SiC doping does not require high-temperature synthesis to obtain the higher $J_c(H)$ values [25].

**(D) Magneto-transport and upper critical field ($H_{c2}$):**

Since the 750°C series of samples show the best values of $J_c(H)$ and $H_{irr}$ therefore, we are presenting the magneto-transport measurement up to 14 Tesla applied field for these samples. Figure 7(a) shows the superconducting transition region of magneto-transport measurement up to 14 Tesla for (I) pure, (II) 5-wt% & (III) 10-wt% n-SiC doped samples. Here, we note that the transition is very sharp at zero field for all the samples but the transition width increases with the increase in field value. At low fields, behavior of pure sample is better than that of doped samples. The transition temperature ($T_c$) (where $\rho \rightarrow 0$) for the pure sample is 38.4 K in zero applied field. For the 10 wt% *n*-SiC added sample $T_c$ decreased to 34.5 K in zero applied field due to B site C substitution. Further, it is noted that the $\rho(T)$ curves for the doped samples shifted with increasing magnetic field much more slowly than the pure one. The $T_c$ value for the pure $MgB_2$ is 7.4 K for 14 Tesla applied field while is 14.8 K for the 10-wt % *n*-SiC doped sample under the same field. Thus, addition of *n*-SiC clearly improves the superconducting performance of bulk $MgB_2$ sample at elevated fields.

A further important point is that the nominal resistivity of these samples is very different, $\rho(40K)$ being 35 µΩ-cm for the undoped sample, 84 µΩ for 5 wt % and 155 µΩ for the 10 wt % *n*-SiC doped sample. It refers that the scattering increases with increasing *n*-SiC content. In Figure 7(b) the temperature dependence of normalized resistivity $R(T)/R(300K)$ is shown for pure, 5-wt% and 10-wt% SiC doped samples. The residual resistivity ratio (RRR = $R_{T300K}/R_{Tonset}$) values for the pure, 5-wt% and 10-wt% SiC doped samples are 3.0, 1.75 and 1.5 respectively. Both C doping (revealed by contraction in *a* parameter and reduction in $T_c$) and the inclusion of $Mg_2Si$ (revealed by XRD) can enhance the electron scattering, and hence the decreased RRR values. Further, the higher values of room temperature resistivity for doped samples indicate that the impurity scattering is stronger due to the Carbon substitution at Boron sites. This is in agreement with previous studies on $MgB_{2-x}C_x$ systems [26,27].

The variation of upper critical field with respect to reduced temperature $H_{c2}(T)$ is shown in Figure 7(c). The upper critical field is determined from the resistive transitions shown in Figure 7(a) using the criterion of 90% of the $\rho_N$ value, where $\rho_N$ is the normal resistivity at about 40 K. Both the doped samples show the higher values of critical field in comparison to the pure sample but 10-wt% *n*-SiC doped sample is having the best values. At about 20 K the $H_{c2}$ value for undoped sample is ~10 Tesla and the same is enhanced to 14 Tesla for both the 10-wt% *n*-SiC doped sample. The carbon substitution into boron site in lattice is responsible for this $H_{c2}$ increase due to the disorder on the lattice site of boron. We argue that *n*-SiC reacting with Mg, releases highly reactive free C on the atomic scale at the same temperature where formation of $MgB_2$ takes place.

Because of the availability of reactive C atom at that time, the C can be easily incorporated into the lattice of $MgB_2$ and substitute into B sites [4].

Some theoretical models are needed to be applied to determine the upper critical field values at low temperatures. The experimental data for the $H_{c2}$ of $MgB_2$ can be described with high accuracy by the given expression called Ginzburg Landau (G-L) equation [28]:

$$H_{c2}(T) = \frac{H_{c2}(0)\theta^{1+\alpha}}{1-(1+\alpha)\omega + \ell\omega^2 + m\omega^3} \quad (1)$$

with $\omega = (1-\theta)\theta^{1+\alpha}$ and $\theta = 1 - \frac{T}{T_c}$

In Figure 7(d), we plot the experimental data of $H_{c2}$ as symbols for the second *series* of samples. We also show the theoretical fitting to these data by above G-L equation (shown by solid curves). As shown in Figure 7(d), there is reasonable agreement between theory and experiment. We got the fitting parameters α = 0.41, ℓ = 2.5 and m = -0.95 for 5-wt% *n*-SiC doped sample; these values are matches well with Askerzade et. al [28]. The small deviation between the data and theory may suggest some additional effective bands, which should be considered in the G-L theory in order to get for the exact temperature dependence of $H_{c2}$. From the fitting, we can clearly see that, initially, the behavior of $H_{c2}$ with $T$ is linear near $T_c$ and extends up to a temperature of ~10 K and after that it saturates in the range 3–10 K. Below 3 K the $H_{c2}$ line has negative curvature. The $H_{c2}(0)$ for 5-wt% *n*-SiC doped sample is found to be about 33.2 Tesla while the same is just nearly 17 Tesla for the pure $MgB_2$ sample. Thus, G-L theory also confirms the enhancement of $H_{c2}$ value. The $H_{c2}(0)$ values determined by us are also in confirmation with other reported literature [11,29,30].

**(E) Structure – microstructure and interface analysis by *High Resolution TEM*:**

We carried out the micro-structural investigations employing high-resolution transmission electron microscopy (*HRTEM*) on the 750°C series. Figure 8 shows the *HRTEM* image of unreacted (bare) *n*-SiC particles. It can be seen that *n*-SiC particles are almost spherical in shape having particle size ~20-50 nm with facetted contours on the surfaces of these nanoparticles. In Figure 8, upper inset shows the lattice scale image of an individual nanoparticle exhibiting the inter-planar spacing of 0.25 nm, which corresponds to {111} plane of a cubic SiC (space group: $F\bar{4}3m$) of lattice parameter $a$=0.43 nm. The lower inset shows the electron diffraction (*ED*) pattern of the unreacted *n*-SiC particles. The Debye rings marked as 1 to 4 correspond to inter-planar spacing of 0.25, 0.22, 0.15 and 0.13 nm for {hkl}: {111}, {200}, {220} and {311} planes, respectively.

Figure 9(a) shows the *HRTEM* micrograph of pure $MgB_2$ and corresponding electron diffraction (*ED*) pattern is shown as right inset of the figure. The Debye rings of ED (right inset of Figure 9(a)) correspond to the inter-planar spacing of 0.27, 0.22, 0.15 and 0.12 nm for {hkl}: {100}, {101}, {110} and {201}, respectively. Left Inset of Figure 9(a) depicts the edge of an individual $MgB_2$ grain covered by a very thin layer (average size ~9 nm), which may be attributed to the formation of MgO during synthesis. However we have not explored further about any growth of such MgO layer in the present work but similar effect was observed in literature [31]. Corresponding to the Figure 9(a) lattice scale image of a single grain of pure $MgB_2$ is shown in Figure 9(b). It is observed that a single grain consists of several sub grains inside it of the size of about 3 to 5 nm. Figure 9(b) clearly reveals the random distribution of such nano-crystallites in different orientation. From this micrograph the inter-planer distances of three crystallites (marked as I, II and III in figure) are calculated as 0.21, 0.15 and 0.26 nm, which correspond to the planes {101}, {110} and {100} respectively of $MgB_2$ crystals.

Figure 10(a) and 10(b) show the bright field and corresponding dark field *HRTEM* images of $MgB_2$+10-wt% *n*-SiC sample synthesized at 750$^o$C respectively. Indeed, these *TEM* micrographs show 5 to 15 nm large round shaped particles of $Mg_2Si$, existing within each $MgB_2$ grain. It is to be noted that the inclusions are smaller than the SiC-particles initially added, indicating that the SiC is dissolved in the parent $MgB_2$ during synthesis. Dou et al. [24,32] also observed that SiC particles could dissolve completely or up to a certain extent into $MgB_2$. The corresponding electron diffraction (*ED*) pattern is shown in inset of Figure 10(a). The Debye rings are marked as 1 to 5 correspond to inter-planar spacing of 0.23, 0.22, 0.15, 0.13 and 0.12 nm for {220} plane of $Mg_2Si$, {101}& {110} of $MgB_2$, {422} plane of $Mg_2Si$ and {201} plane of $MgB_2$, respectively. This clearly confirms the presence of $Mg_2Si$ particles within the matrix of parent $MgB_2$.

Another micrograph (Figure 10(c)) exhibits a cluster of $Mg_2Si$ nanoparticles in the matrix phase of $MgB_2$ grain. This image elucidates a large number of dislocations / defects at the interface of $MgB_2$ with the cluster of $Mg_2Si$. Dislocations are known to serve as strong pinning centers along with the *nano* inclusions [33]. Lattice scale image of $MgB_2$+10-wt% *n*-SiC in Figure 10(d) clearly reveals the presence of $Mg_2Si$ dispersed in the $MgB_2$ matrix. A white line boundary is marked around the $Mg_2Si$ particle in the figure. The particle size of $Mg_2Si$ is nearly 9 nm. In this micrograph {220} plane of $Mg_2Si$ is being visualized with inter-planar distance 0.23 nm.

At some instances the un-reacted (un-dissolved) traces of SiC are also found visible during the *HRTEM* examination of these samples. These SiC particles are forming a clear interface with the matrix phase of $MgB_2$. As an illustrative example, Figure 11(a) and 11(b) represents lattice scale image of the $MgB_2$+10-wt% *n*-SiC sample and corresponding *EDS* (Energy Dispersive Spectrum) pattern, respectively. Lattice planes of $MgB_2$ and SiC particle (which is embedded in parent $MgB_2$ matrix) can be clearly resolved in Figure 11(a). The inter-planar distances are marked as 0.26 nm and 0.22 nm respectively for $MgB_2$ and SiC in the image, which corresponds to the plane {100} of $MgB_2$ and {200} of SiC respectively. A corresponding *EDS* analysis of the microstructure delineated the presence of Mg, B, C, Si, and O peaks in the sample.

This, and the results of XRD, infers that the inclusions of *nano* particles were primarily made of $Mg_2Si$ and with the traces of un-reacted SiC. Therefore, it can be concluded that first reaction of *n*-SiC with Mg forming $Mg_2Si$ and second free C being incorporated into $MgB_2$ both helps in pinning of vortices and improved superconducting performance. $Mg_2Si$ and excess carbon can be embedded within the grain of $MgB_2$ as *nano* inclusions. Must be noted that this nano inclusions with dimension smaller than 10 nm are of the right dimension to fit with the coherence length in Magnesium Diboride and, therefore, are suitable to act as additional pinning center. In Neutron irradiated polycrystalline samples the correlation between defects on nano-metric scale observed by TEM and $J_c$ enhancement was clearly demonstrated. [34]; here we suggest that the $J_c$ enhancement at high field can be due a similar effect. Naturally the better $J_c$ in-field behavior takes advantage of the increased critical field. Our results suggest that a combination of substitution-induced defects and highly dispersed additives are responsible for the enhanced flux pinning in the doped samples.

## Conclusions

In summary, a systematic study on the effect of synthesis temperature on the phase formation, lattice parameters, *FWHM*, actual C content, critical current density and irreversibility field of *n*-SiC doped MgB$_2$ is presented in this article. We observed that optimal doping effect of *n*-SiC was achieved for synthesis temperature of 750$^o$C. We can say that two factors are responsible for the best performance of presently studied and relatively low temperature (750 $^0$C) synthesized *n*-SiC doped MgB$_2$, i.e., (a) C substitution for B induces disorder in lattice sites with increased $H_{c2}$, and (b) the reduction in grain size, extra defects and embedded inclusions such as Mg$_2$Si and unreacted *n*-SiC enhances the flux pinning. To know the exact effect of increased $H_{c2}$ or the pinning induced defects etc on the $J_c(H)$ in general , one plots the flux pinning against the reduced field (H/H$_{irr}$) [3,32-34]. Unfortunately, in presently studied higher $J_c$ samples of MgB$_2$ being doped with *nano*-SiC and processed at low temperature, the avalanche of flux jumps below 20kOe do not permit clear observation of the flux pinning peak and hence meaningful conclusion on the nature of pinning can not be drawn. It is clear from Fig. 6(a) that $J_c(H)$ performance of 750 $^0$C synthesized MgB$_2$ is significantly improved with *n*-SiC doped inclusions. The $T_c(R=0)$ of above 15K for the studied *n*-SiC doped and 750 $^0$C synthesized MgB$_2$ under 140 KOe field is one of the highest values yet obtained for variously processed and *nano*-particle added MgB$_2$ in literature to our knowledge.

## Acknowledgement

Authors from NPL are thankful to Prof. R.C. Budhani (*DNPL*) for his keen interest in present work. Arpita Vajpayee would like to thank CSIR for providing the SRF scholarship for her Ph.D. work.

Figure and Table Captions:

Figure 1(a): Powder XRD patterns for 5-wt% & 10-wt% *n*-SiC doped samples along with pristine MgB$_2$ sample synthesized at 700°C, 750°C, 800°C & 850°C

Figure 1(b): Variation of *a*-axis lattice parameter, c/a, actual C substitution level (x) in Mg(B$_{1-x}$C$_x$)$_2$ and *FWHM* of (110) peak with synthesis temperature.

Figure 2: (a) DC susceptibility in FC & ZFC case and (b) AC susceptibility for pure, 5-wt% & 10-wt% *n*-SiC doped MgB$_2$ samples prepared at 750°C

Figure 3: Dependence of transition temperature ($T_c$) on synthesis temperature ($T_s$) for pure, 5-wt% & 10-wt% *n*-SiC doped MgB$_2$ samples

Figure 4: Magnetization loop *M(H)* for MgB$_2$+*n*-SiC$_x$ (x=0%, 5-wt% & 10-wt%) samples of *second series* (Synthesized at 750°C) up to 14 Tesla field at 5, 10 & 20 K

Figure 5(a): Effect of synthesis temperature ($T_s$) on irreversibility field ($H_{irr}$) for pure, 5-wt% & 10-wt% doped MgB2 samples at 5, 10 & 20 K

Figure 5(b): Variation of irreversibility field ($H_{irr}$) with respect to *n*-SiC concentration at 5, 10 & 20 K for $T_s$=750°C

Figure 6(a): The critical current density versus magnetic field $J_c(H)$ plots at 20 & 5 K for SiC doped and undoped samples synthesized at 700, 750, 800 & 850°C

Figure 6(b): Dependence of $J_c$ on synthesis temperature at 5 K and 6 Tesla for pure MgB$_2$ and at 5 K and 9 Tesla for doped samples

Figure 7(a): Superconducting transition zone of resistivity vs temperature plot under applied magnetic field ρ(*T*)H up to 14 Tesla for (I) pure, (II) 5-wt% & (III) 10-wt% *n*-SiC doped samples synthesized at 750°C

Figure 7(b): Variation of resistivity with temperature ρ($T$) for Pure, 5-wt% and 10-wt% *n*-SiC doped samples synthesized at 750°C

Figure 7(c): Upper critical field ($H_{c2}$) vs reduced temperature ($T/T_c$) plot for Pure, 5-wt% and 10-wt% *n*-SiC doped samples synthesized at 750°C

Figure 7(d): Theoretical fitting by G-L equation of upper critical field ($H_{c2}$) values for undoped, 5-wt% and 10-wt% *n*-SiC doped samples of *second series*. The symbols denote the experimental data and solid lines are the best-fit curve according to G-L equation.

Figure 8: *HRTEM* micrographs of *n*-SiC particles; upper inset shows the lattice scale image of an individual nano particle and lower inset shows the electron diffraction (*ED*) pattern of the same.

Figure 9(a): *HRTEM* micrograph of pure $MgB_2$ samples of *second series* ($T_s$=750°C); Right inset shows the corresponding electron diffraction (*ED*) pattern and left inset shows an individual $MgB_2$ grain covered by a thin layer of MgO.

Figure 9(b): Lattice scale image of pure $MgB_2$ samples of *second series* ($T_s$=750°C) showing the presence of different *nano* crystals.

Figure 10(a): Bright field *HRTEM* images of $MgB_2$+10-wt% *n*-SiC sample synthesized at 750°C; inset shows the *ED* pattern of the same.

Figure 10(b): Dark field image of the $MgB_2$+10-wt% *n*-SiC sample synthesized at 750°C; Arrows show the presence of few $Mg_2Si$ particles in the sample.

Figure 10(c): Dark field image of the $MgB_2$+10-wt% *n*-SiC sample synthesized at 750°C.

Figure 10(d): Lattice scale image of a grain of $MgB_2$+10-wt% *n*-SiC sample synthesized at 750°C. A boundary is drawn around an $Mg_2Si$ particle.

Figure 11(a): Lattice scale image of an interface between $MgB_2$ and SiC from a $MgB_2$+10-wt% *n*-SiC sample synthesized at 750°C.

Figure 11(b): Corresponding *EDS* (Energy Dispersive Spectrum) pattern of the $MgB_2$+10-wt% *n*-SiC sample synthesized at 750°C.

Table 1: Lattice parameters 'a' and 'c' calculated from the peak shifts in XRD patterns for undoped and SiC doped samples of all the four *series* synthesized at different temperatures

Table 1:

| Sample Name | $T_s$ = 700 °C | | $T_s$ = 750 °C | | $T_s$ = 800 °C | | $T_s$ = 850 °C | |
|---|---|---|---|---|---|---|---|---|
| | 'a' (Å) | 'c' (Å) | 'a' (Å) | 'c' (Å) | 'a' (Å) | 'c' (Å) | 'a' (Å) | 'c' (Å) |
| Pure $MgB_2$ | 3.0844(92) | 3.5208(40) | 3.0829(06) | 3.5203(83) | 3.0803(47) | 3.5191(89) | 3.0854(58) | 3.5253(91) |
| $MgB_2$+5% *n*-SiC | 3.0844(13) | 3.5210(66) | 3.0774(54) | 3.5259(60) | 3.0712(60) | 3.5176(88) | 3.0796(91) | 3.5293(83) |
| $MgB_2$+10% *n*-SiC | 3.0799(42) | 3.5227(44) | 3.0685(26) | 3.5209(52) | 3.0672(17) | 3.5206(93) | 3.0720(07) | 3.5299(99) |

Figure 1(a):

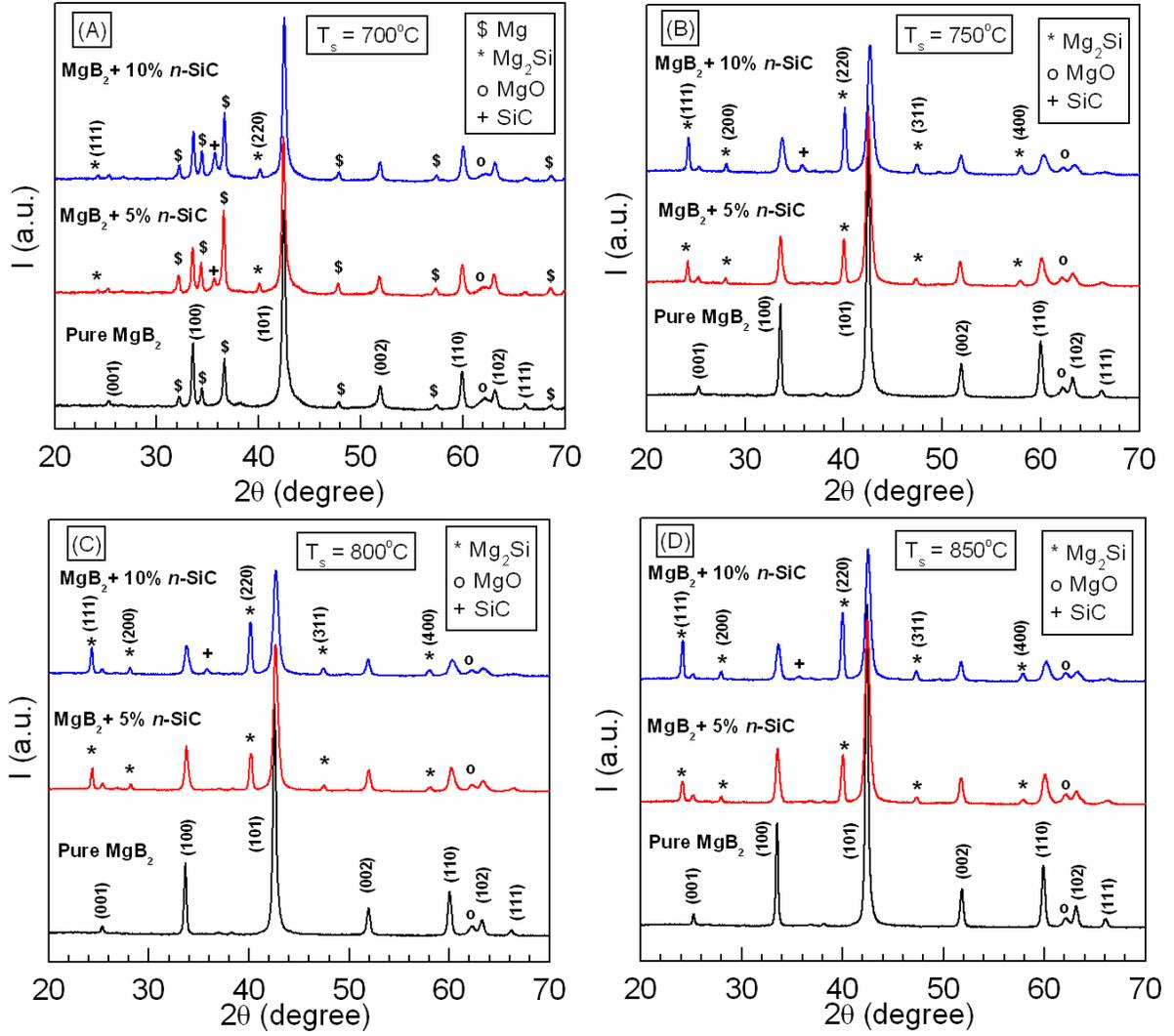

Figure 1(b):

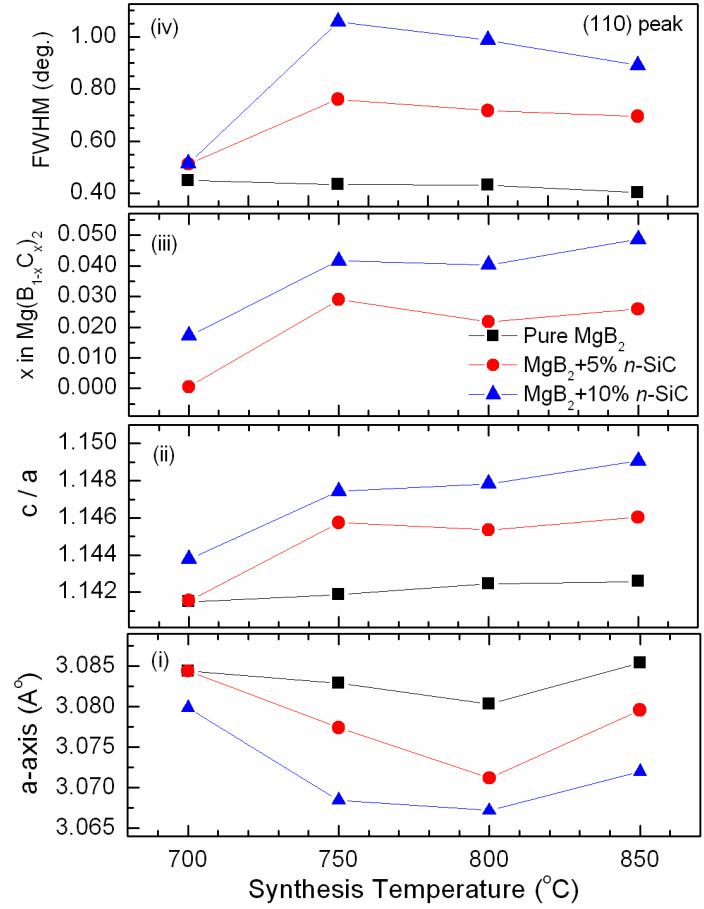

Figure 2:

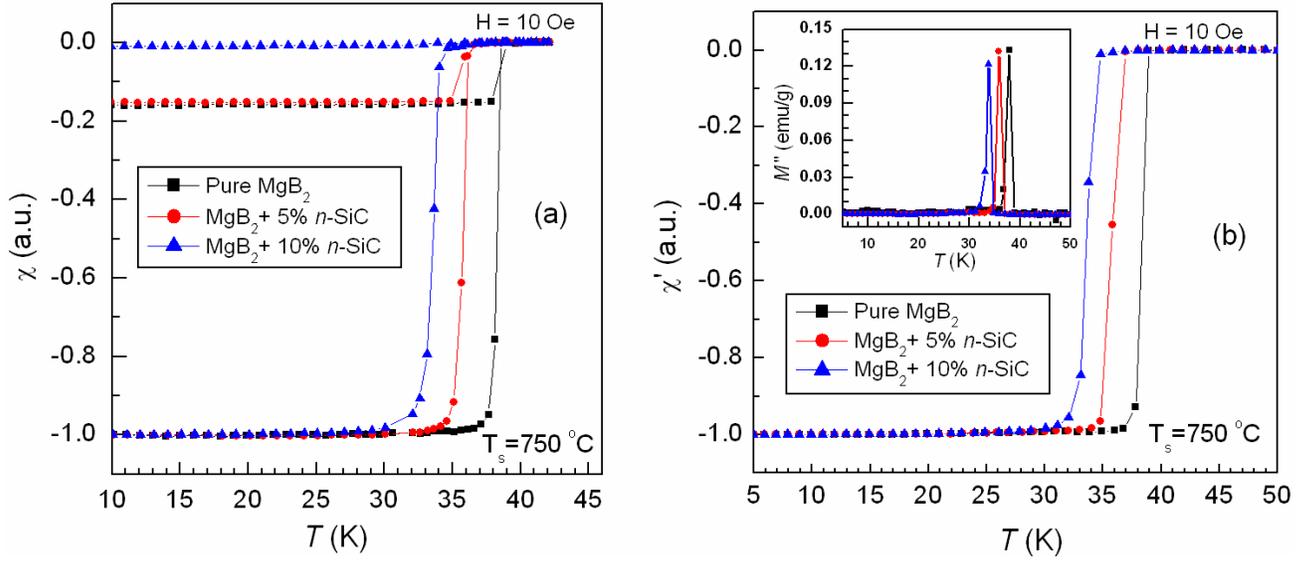

Figure 3:

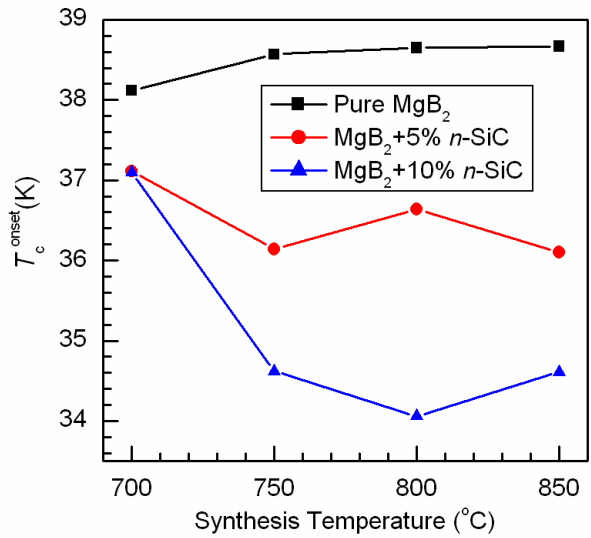

Figure 4:

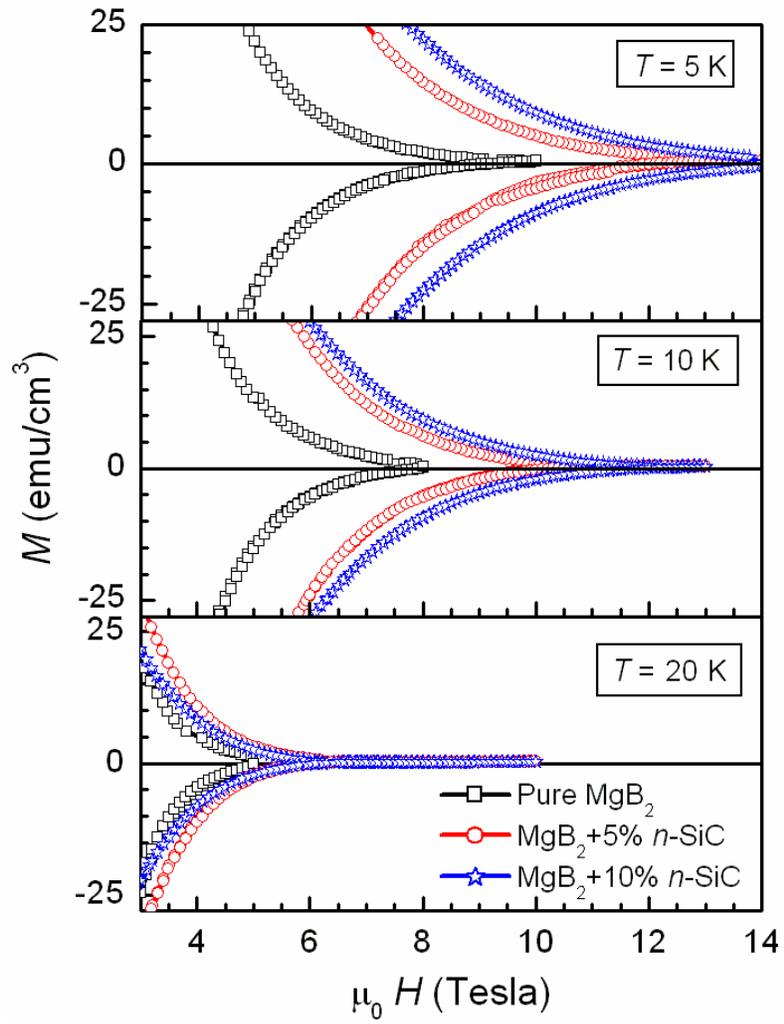

Figure 5(a):

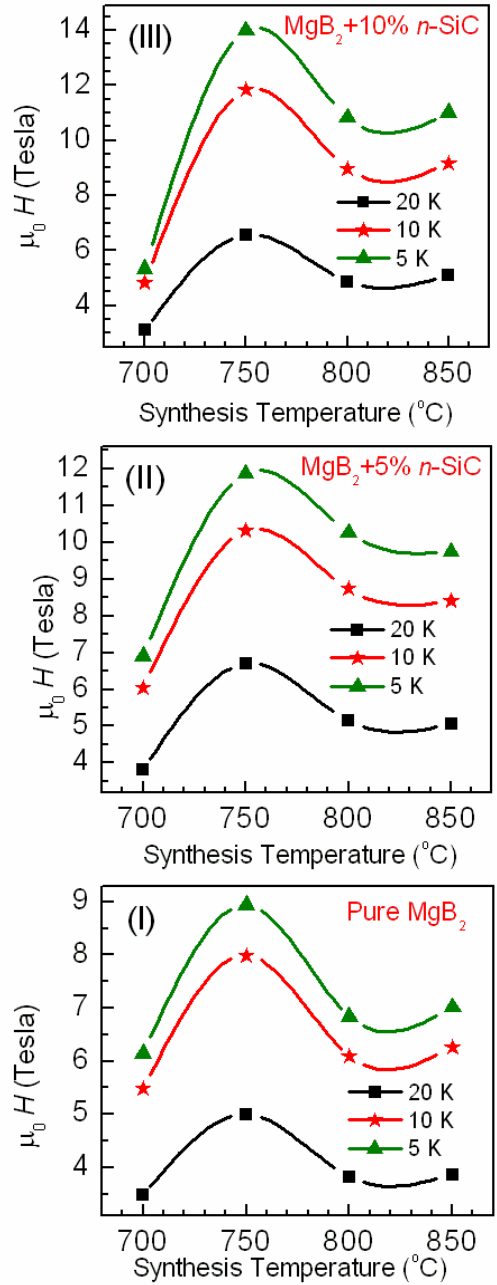

Figure 5(b):

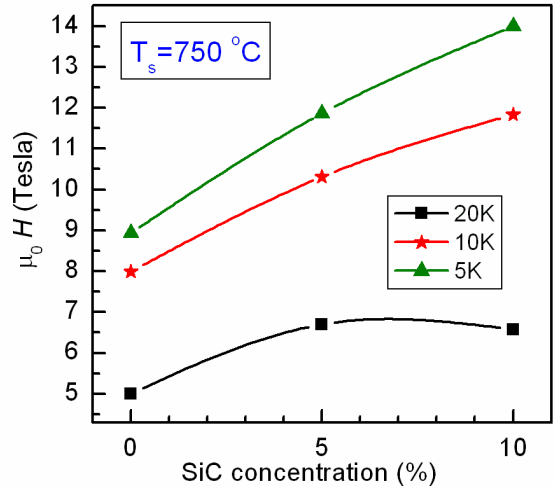

Figure 6 (a):

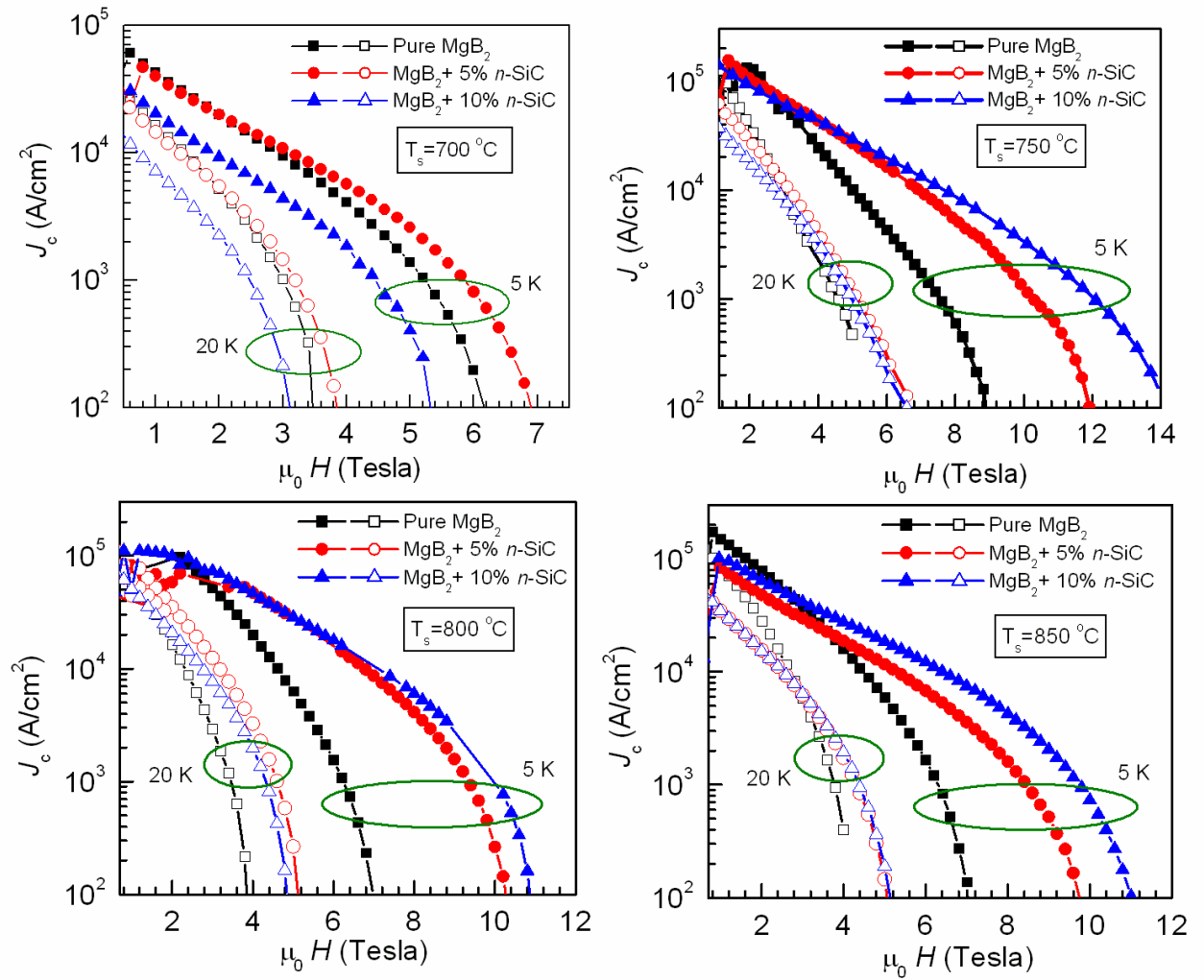

Figure 6 (b):

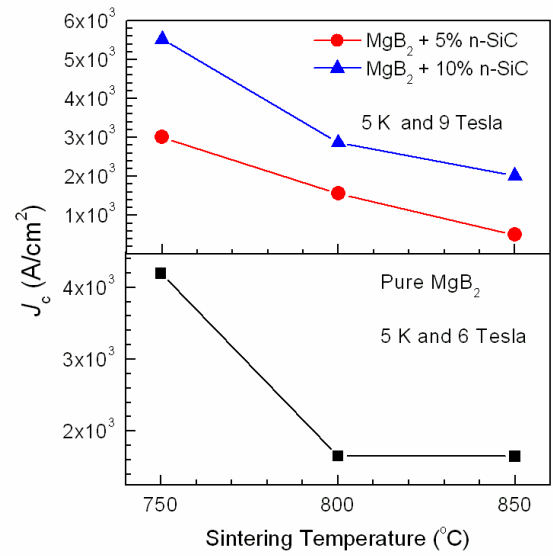

Figure 7(a):

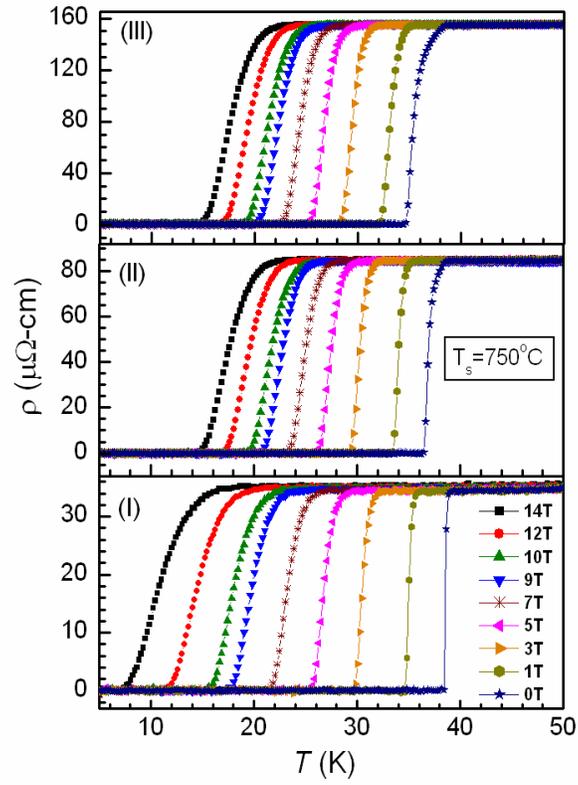

Figure 7(b):

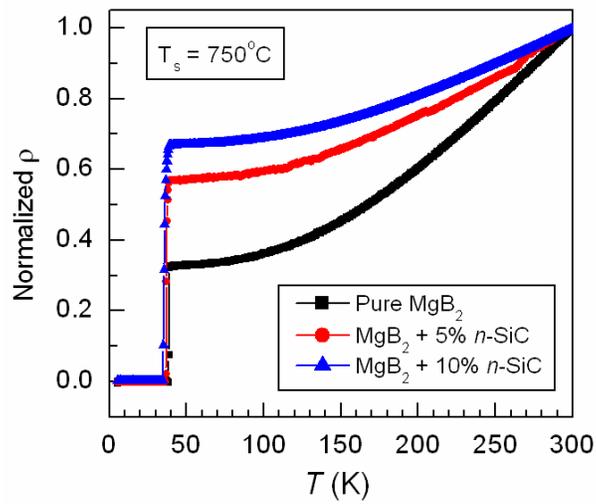

Figure 7(c):

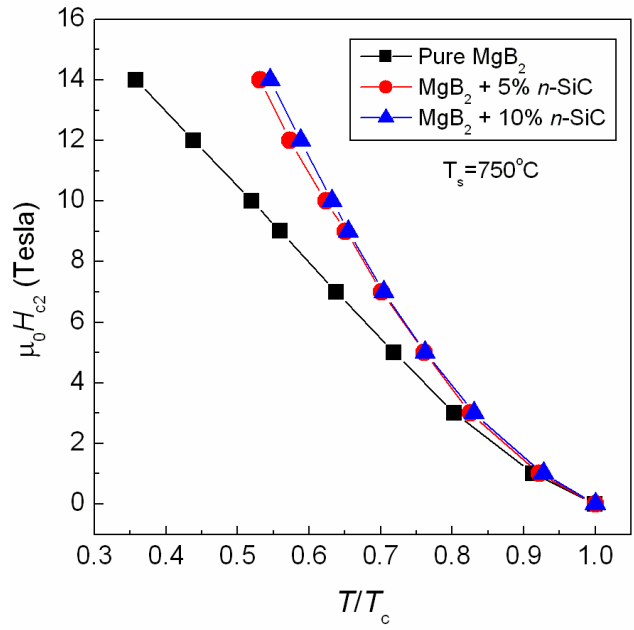

Figure 7(d):

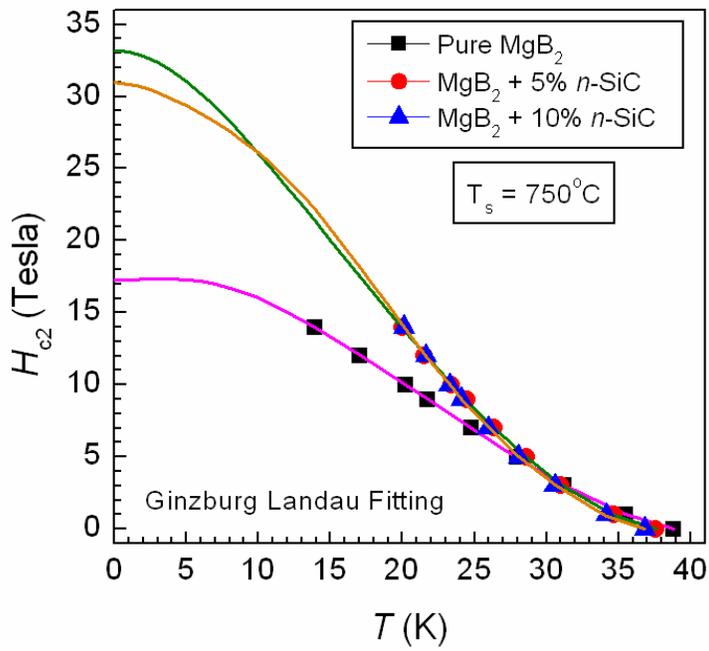

Figure 8:

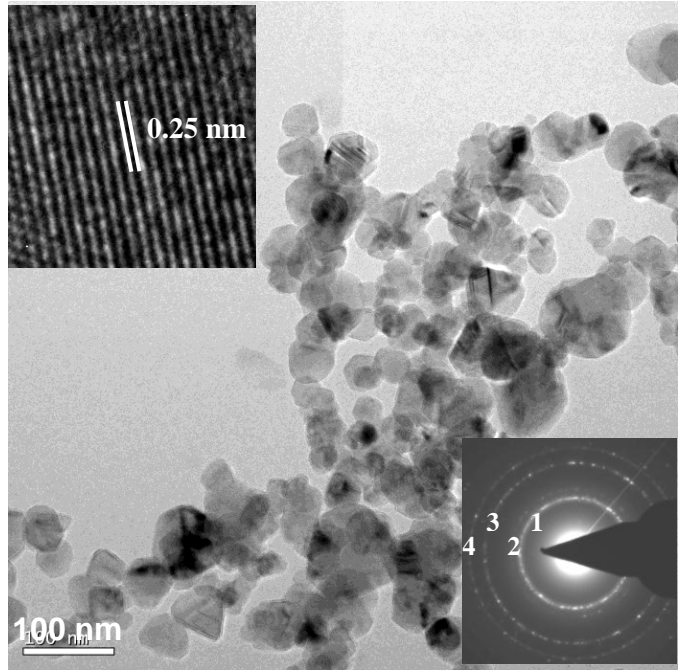

Figure 9(a):

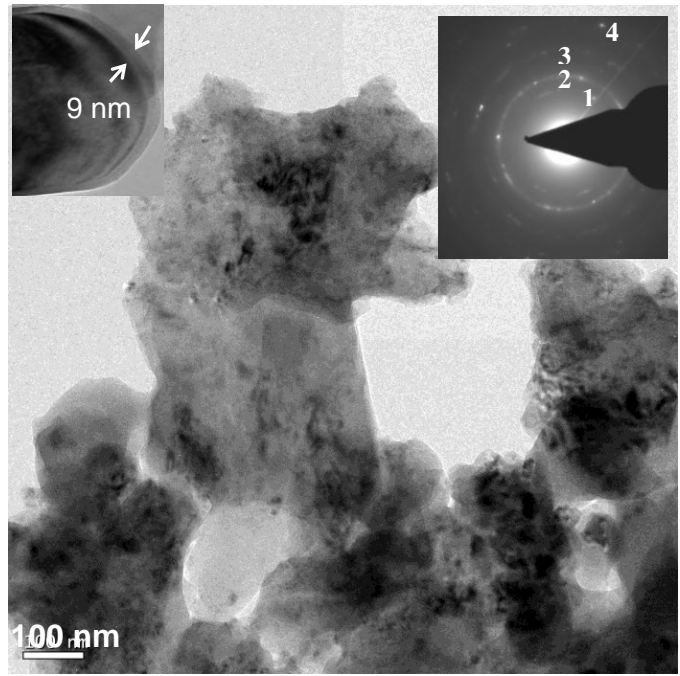

Figure 9(b):

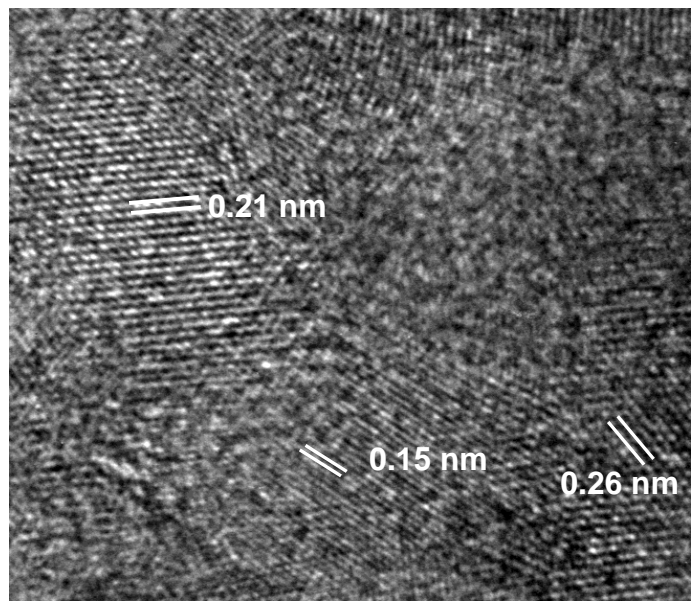

Figure 10(a):

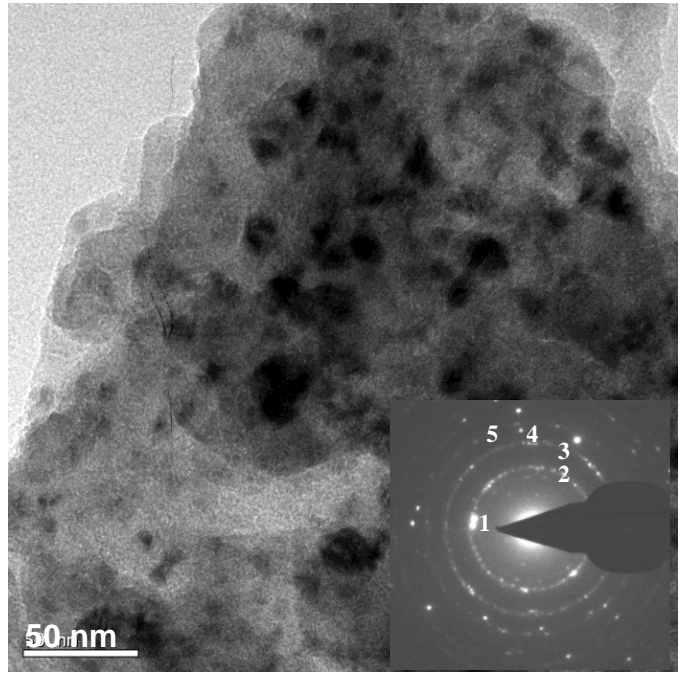

Figure 10(b):

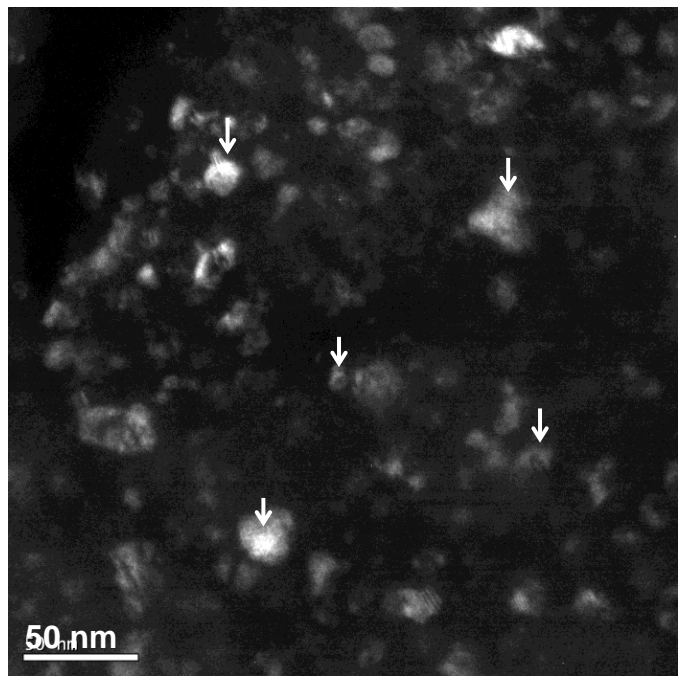

Figure 10(c):

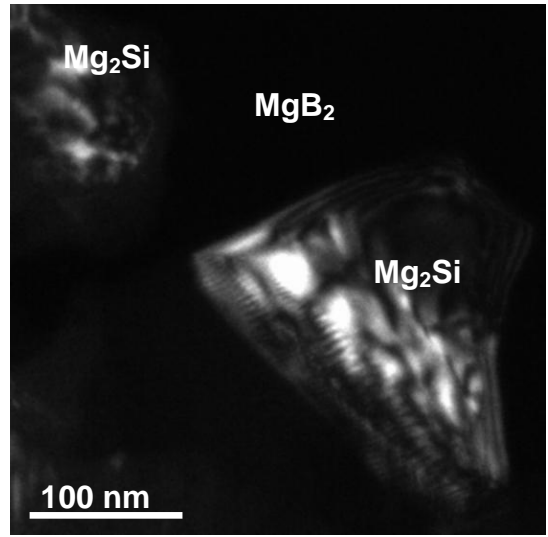

Figure 10(d):

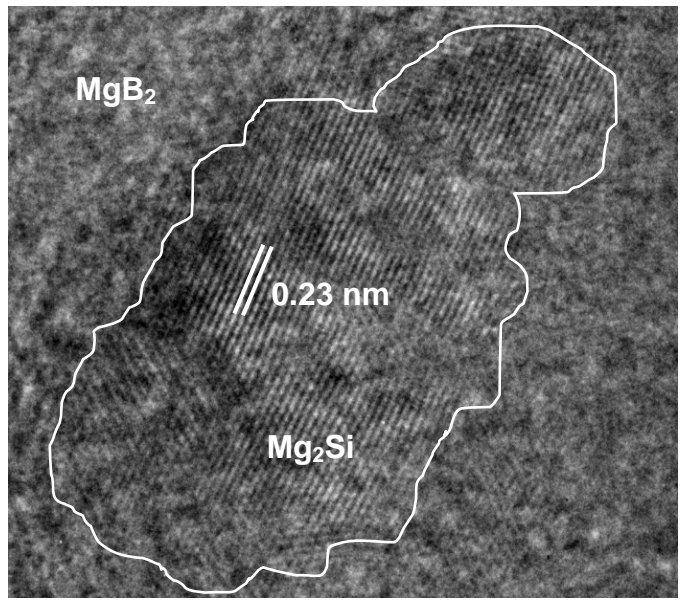

Figure 11(a):

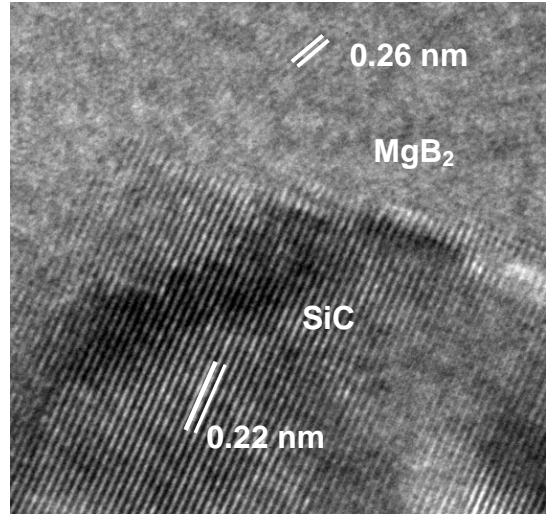

Figure 11(b):

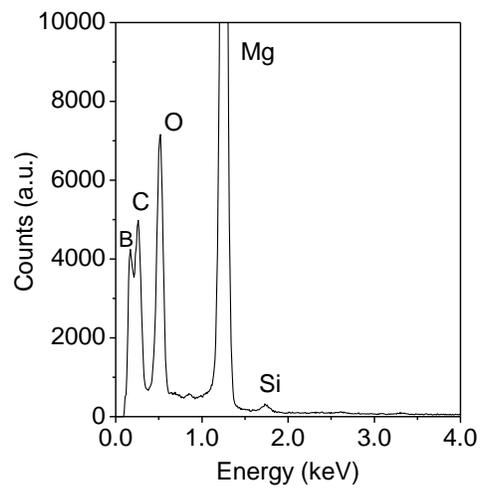